# Towards A Unified Utilitarian Ethics Framework for Healthcare Artificial Intelligence

Forhan Bin Emdad[1], Shuyuan Mary Ho[1], Benhur Ravuri[1], and Shezin Hussain[1]

1 College of Communication & Information, Florida State University, Tallahassee, United States

**Abstract**

Artificial Intelligence (AI) aims to elevate healthcare to a pinnacle by aiding clinical decision support. Overcoming the challenges related to the design of ethical AI will enable clinicians, physicians, healthcare professionals, and other stakeholders to use and trust AI in healthcare settings. This study attempts to identify the major ethical principles influencing the utility performance of AI at different technological levels such as data access, algorithms, and systems through a thematic analysis. We observed that justice, privacy, bias, lack of regulations, risks, and interpretability are the most important principles to consider for ethical AI. This data-driven study has analyzed secondary survey data from the Pew Research Center (2020) of 36 AI experts to categorize the top ethical principles of AI design. To resolve the ethical issues identified by the meta-analysis and domain experts, we propose a new utilitarian ethics-based theoretical framework for designing ethical AI for the healthcare domain.

**Introduction**

Artificial Intelligence (AI) has revolutionized the healthcare sector. Particularly, different AI techniques such as machine learning (ML) and deep learning (DL) have played a vital role in clinical decision support for diagnosis, prognosis, text and image classification, and treatment. AI-based health outcome prediction has contributed largely to the mortality prediction of patients with diseases such as heart disease, cancer, stroke [1], and Alzheimer's disease. The recent trend of the use of transformer-based large language model AI tools such as ChatGPT (GPT-3), DALL-E, etc. can highly contribute to the improvement of the healthcare quality outcome. Although, this trend brings the question of the ethical use of AI in different sensitive domains such as education and healthcare. AI can be weaponized to create chaotic situations. For example, students can use AI to plagiarize their thesis study. Similarly, healthcare data can be manipulated with AI to generate erroneous predictions. To minimize the ethical concerns about using AI, a unified ethical framework is required which will provide some guidelines to the AI practitioners and create safer AI for the users. Despite having many existing ethical frameworks, there is a rising concern over the process of implementing a unified ethical framework comprising the influential factors related to the healthcare domain.

In the healthcare context, clinicians still restrict themselves from trusting AI for clinical decision support. Most clinicians have raised concerns over the use of AI in healthcare because of a few major challenges, such as the opaqueness of algorithms, inconsistent data, safety and transparency, and algorithmic biases. This problem was even observed in Google DeepMind, which brought revolution and was accused of concerns such as lack of transparency leading to a lack of trust among clinicians. These major challenges can be mapped into ethical and legal challenges. According to most studies, transparency, fairness, and privacy issues fall in the category of ethical issues [2].

By mitigating the challenges of AI, it would be easier to design ethical AI in critical sectors like healthcare. There are existing ethical approaches describing how ethics can be incorporated into the medical domain and AI. Examining the relevant ethical approaches, we can identify generalized constructs related to ethical AI design. By definition, AI is a field of computer science that conducts different simulations to act like human intelligent systems by learning from a huge amount of data. Therefore, as humans require moral or ethical guidance to become good human beings, trustworthy ethical AI can be designed by following a well-established ethics framework. Specifically, successful ethical AI design in healthcare would require the identification of the issues related to AI supported by empirical evidence. To attain the full potentiality of designing an ethical AI framework study, the perception of developers, ethicists, healthcare experts, and other stakeholders should be taken into consideration.

In reviewing the many ethics frameworks, we adopt the utilitarian's ethics perspectives in designing ethical AI in healthcare. We investigate an important research question: "What ethical factors can influence the ethical AI design in healthcare?" We conduct a qualitative study to analyze the expert's opinions from the healthcare industry from a survey study conducted by PEW Research Center (2020). Critical concepts were extracted and analyzed to confirm the principal ethics issues in healthcare. The paper describes a conceptual ethics framework based on utilitarian's perspectives.

**Why Ethics**

Ethics can be simply defined as how the world should act properly. However, it becomes really challenging to define ethics in different contexts such as academic or healthcare or specifically AI as a system because there are no descriptive dimensions for ethics. Ethics can be achieved by feelings or habits, religious beliefs, government law, or cultural norms. These ethics are not defined descriptively. Therefore, we can try to look at some of the common approaches to understanding ethics, then we can try to fit perspective lenses into the healthcare AI domain.

However, there are problems raised by ethical challenges which worry the AI users. These concerns are related to the use of AI in society, industry, the military, and healthcare. Moreover, ethical concerns become vigorous when unprecedented incidents can take place due to the use of AI and there will be no one to be held accountable as AI cannot be accountable for the incidents. Extreme cases of these incidents would be if AI kills any civilians in war or deaths happen due to the wrong use of AI in healthcare. Maintaining ethical principles in the use of AI would be helpful to minimize the risk. Further, the need for ethics can be demonstrated by the below reasons:

- **Bigdata.** Businesses and organizations are collecting big data**.** Collecting big data is problematic because big data includes privacy and personal data. System developers should use the data ethically as data should be used for the core purpose which is good for society.

- **Misuse of Data.** Data can be manipulated. Manipulated data can be made biased, and can be used to build algorithms to serve an unethical purpose.

- **Humans or Bots?** The recent advent of chatbots and voice-related AI assistance are hard to differentiate from humans. Chatbots are used in healthcare dedicated to mental health conditions. Though, there comes the issue of the acceptability of health chatbots among patients without proper consent. The ethical use of AI is undoubtedly needed to prevent this type of fraud.
- **Against good AI society**. The major objective of AI is to only benefit society, not to harm society[3]. To bring this objective to reality, we need ethics implemented in building AI.

In addition, few researchers have categorized ethical challenges in different levels to propose a real-world ethical framework. Goirand et al., (2021) [4] divided ethical challenges into ethical principle level, design level, technology level, organizational level, and Regulatory level. However, researchers are delving more into the technological level to derive insights about implementing a robust ethical framework. In our study, we subdivided the technology level into other levels for further analysis.

**Utilitarian Ethics**

To understand utilitarian ethics, first, we must observe ethics through major philosophical lenses. There are four orientations of the ethics lenses that are mostly adopted: Kantianism or deontology, utilitarianism or consequentialism, contractarianism, and Virtue Ethics. First, Kantianism understands ethics as a moral law-adhering concept. Kantianism is also known as deontology. Kantianism is the most popular ethical lens among health AI researchers because this approach helps them understand the how system was developed and what rules were followed while designing it. According to Kantianism, an individual's responsibility is to find out the moral law for themselves which will be rational. Kantian ethical theory is widely used in medical ethics. Although, as Kantian ethics depends much on the laws, critics argue that it cannot resolve real-world medical complexities and will lose its focus on autonomy. However, the question will arise whether the ethics-related laws are correct or can they be minimized, or what will be the consequences of removing the ethical laws or what are the most important laws among them.

On the other hand, utilitarianism ethics mentions that AI actions should be for the greater good or happiness which is measured with utility. Utilitarianism is also known as consequentialism meaning the consequence of the action should be for happiness. In social science and computer science, utilitarianism is mostly used. In utilitarianism, "utility" is the representation of individual good. Societal good comes from the sum of individual utilities. Even game theory can be defined with utilitarianism ethics theory as it provides a reward to individuals with higher utility where the utility can be quantitatively measurable. Although, it is argued that utility should not be the only measure in the healthcare domain.

Another famous and old concept-based theory is the social contract theory. Social contract theory is the philosophical view that an individual has to obey the restrictions created by social contracts. These contracts act like a law for the individuals when they violate these laws they get punishment or send to prison. These contracts are made on the mutual agreement of the society members based on moral and political behavior. Although in the context of healthcare AI, contractarian ethics (based on social-contract theory) is criticized as it does not address health disparities, and health disparities (algorithmic bias in healthcare) increase morally problematic injustices.

Contrary to Kantianism, virtue ethics states that ethics should be developed by virtues and living experiences instead of set laws. The virtue ethics approach is widely used in AI and society. Virtue ethics can differ from culture to culture as there is no universal habit or morality. It is believed that AI might replace humans. However, according to the virtue ethics approach, clinicians will play a vital role in designing and regulating virtuous AI. In addition, the challenge remains such fair and virtuous AI will be hard to build as AI forms its habit from the data it has gathered.

Among the other ethical lenses, utilitarian ethics is the most suited approach for evaluating healthcare AI as a greater number or quantity of utility helps users to understand clearly which works better for their happiness or satisfaction. Utilitarianism believes that good actions with the correct guidance will bring greater satisfaction and trust, otherwise, they will bring unhappiness or suffering. Moreover, the utilitarian approach in the use of AI aligns with bioethical principles such as autonomy, non-maleficence, beneficence, and justice[5]. The utilitarian principle supports the concept that healthcare should not be focused on specific ethnicities or cultures. Utilitarianism also adopted the non-maleficence principle as it considers the greater good[6]. Similarly, utilitarianism relies on higher utility for the benefit of the users. The utilitarianism approach determines any interventions fairly and equitably to follow the justice principle of bioethics.

**Principal Ethics in Healthcare**

Many kinds of research have been conducted to identify the critical ethical challenges and principles in healthcare AI. Čartolovni et al. (2022) have indicated the most important issues related to the recent developments of AI in healthcare considering the ethical aspects, are patient safety, algorithms, transparency, bias, explainability, trustworthiness, opacity, validity and reliability, liability, and accountability [7]. Similarly, another study indicated that the most important concerns are biases, lack of transparency, and privacy concerns of AI systems in healthcare. However, most studies pointed out the lack of regulation as a major ethical issue of AI in healthcare.

In recent times, many reporting guidelines have been published for ethics-based oversight of AI in healthcare, which reflects a strong commitment to transparency and fairness of AI algorithms. The features mentioned in these reporting guidelines provide appropriate measures to evaluate the design of the AI. AI reporting guidelines including MINIMAR (Minimum Information for Medical AI Reporting), CONSORT AI (Consolidated Standards of Reporting Trials), SPIRIT AI (Standard Protocol Items: Recommendations for Interventional Trials-Artificial Intelligence), and FUTURE AI (consist of (i) Fairness, (ii) Universality, (iii) Traceability, (iv) Usability, (v) Robustness, and (vi) Explainability) are considered as most useful for different areas of medical reporting such as diagnosis, prognosis, and medical imaging.

| Principles | | Guidelines | | | |
|---|---|---|---|---|---|
| **Names** | **Key Constructs** | MINIMAR | CONSORT AI | SPIRIT AI | FUTURE AI |
| Bioethical principles | Autonomy, Non-maleficence, Beneficence, and Justice | √ | √ | √ | √ |
| Healthcare AI regulations authorized by European Union (EU) Commission | Oversight | | | | |
| | Technical Robustness and Safety | | √ | √ | √ |
| | Privacy | | | | |
| | Transparency | √ | √ | √ | √ |
| | Fairness | √ | √ | √ | √ |
| | Societal Well-being | | | | |
| | Accountability | | | | |

**Table 1. Ethics dimensionality and Guidelines from the literature**

Many researchers proposed a unified framework containing all the basic constructs of ethics (also can be referred to as bioethics principles) such as beneficence, non-maleficence, autonomy, and justice. AI4people is considered one of the useful ethical frameworks for a good AI society (Floridi et al., 2018). Merely implementing ethical principles will not be sufficient to acquire the best results in building ethical AI, even continuous auditing or oversight is also an important part of the development of AI. Moreover, ethics-based auditing or oversight processes can lead to building trustworthy AI [8]. European Union (EU) researchers also indicated that oversight is a major principle for developing trustworthy AI.

**Method**

We obtained a copy of the survey study conducted by the Pew Research Center (2020). This survey study included surveys and interviews with 36 AI experts from June 30 to July 27, 2020 [9]. We analyzed the data based on the coding scheme (table 2). This section describes the data analysis procedures from our study conducted on the data collected from the survey on the domain experts considering ethical AI design. This section also includes the identification of major ethical challenges and principles of healthcare AI.

*Data analysis and procedure*

The data was coded by 3 coders. The coding process was conducted in three phases. In the first phase, the coding was performed by the coders individually. Then, all the codes were collected and distributed among the coders. In the second phase, all three coders gathered to share their views and justify their coding. In the final phase, the coders produced another updated and robust coding taxonomy. Figure 1 shows the overall workflow of our study. Firstly, we extracted data related to healthcare from the Pew research report. Next, the survey data was imported into the qualitative data analysis tool, and themes were coded, queried, and visualized. Finally, data were analyzed and interpreted to explain the findings in an understandable way.

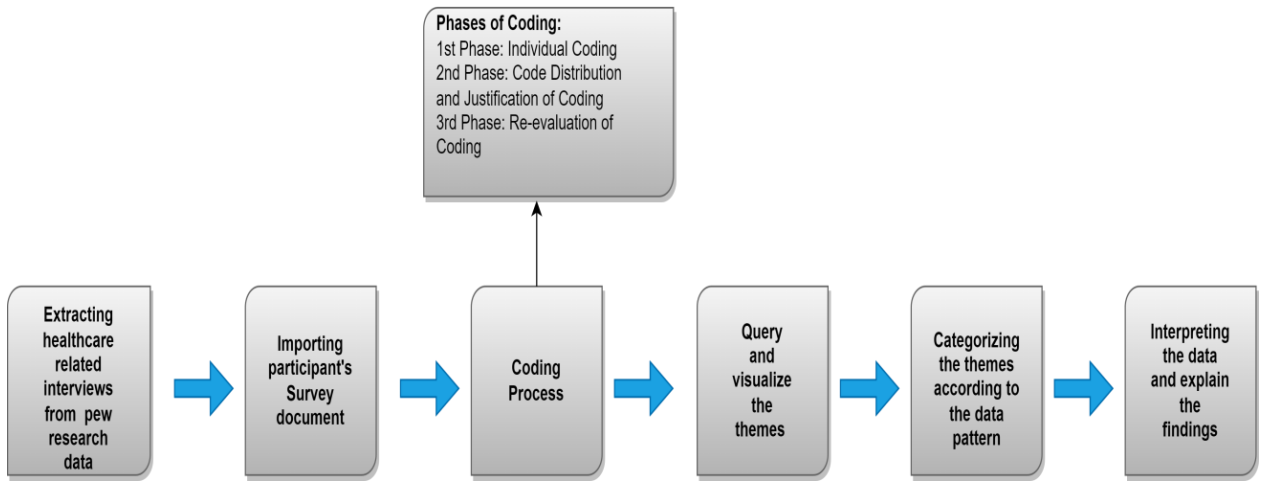

**Figure 1. Overall Workflow of This Study**

The research team processed the survey data that contained information related to AI in healthcare. Specifically, the search terms "healthcare" and "medical" were applied to identify the relevant data to process. The survey data that contain the words "healthcare" and "medial" were extracted from the survey data for coding and analysis. Selected experts were either scholars or higher officials in different organizations dealing with AI. Experts were asked to answer the following questions in the survey: Will AI mostly be used in ethical or questionable ways in the next decade? What gives you the most hope? What worries you the most?

*Data Analysis Tool*

We analyzed a total of 700 lines of data from 36 files. The themes were categorized according to data patterns. NVivo 12 Plus tool was used to format the codes and keep track of the themes, oversight the word cloud, and observe the frequency of the themes. We had a team of three researchers who coded the data separately and conducted the thematic analysis. Researchers compared their analysis and merged their work in NVivo 12 Plus tool.

*Coding Process and Validation*

Nodes were generated from the extracted data by 3 coders: Coder1, Coder2, and Coder3. In this study, we evaluated the inter-rater reliability (IRR) for validating the method of this study. IRR was developed by Miles and Huberman [10], which calculates with the below formula:

IRR = number of agreements between coders/ All codes from the coder *100

IRR measures the consistency of coding between the coders. There are some other methods for these measurements such as Cohen's Kappa, Scott's Pi, or Krippendorff's Alpha. However, IRR is known as the most popular measurement used for qualitative research. In our study, we found 9 common nodes or agreements. Coder1 has 12 total codes with IRR of 75%, Coder 2 has 15 codes with IRR of 60%, and Coder 3 have 14 codes with IRR of 64%. 75% IRR is considered an acceptable score. Therefore, our coding analysis is consistent.

Table 2 in the next page provides a detailed description of the codes agreed by the three coders, frequency of the codes generated from the coder with highest IRR score, categorization of the codes in the different technology sublevel, and definitions of the cod

| Codes Agreed by 3 Coders | Frequency | Level of Analysis | Definition |
|---|---|---|---|
| Accountability/ Responsibility* | 3 | Systems; Data Access; | **Accountability/ liability/ responsibility** refers to the identification of the AI system or individual or developer answerable for the defective activity in the AI designing process |
| Beneficence* | 4 | Algorithms; Systems | The **beneficence** principle of AI refers to the main objective of AI which should be focused towards the greater good of humanity |
| Bias * | 9 | Algorithms | **Bias** refers to the missing values or small sample of underrepresented minorities such as ethnicity or race |
| Explainability/ interpretability* | 1 | Algorithms | **Interpretability** of AI presents the meaning of the algorithm and visualizes the feature's importance |
| Justice & Solidarity * | 5 | Algorithms; Data access; | **Justice** is the principle which allows AI to treat every individual equally in terms of algorithms or data |
| Lack of regulations* | 12 | Policy; Organizational; | **Lack of regulation** refers to the monitoring of ethical and regulatory aspects of the application of AI in healthcare |
| Reliability* | 3 | Systems; | **Reliability** refers to the consistency of the measure meaning if the model generates the same result when it is executed |
| Risks* | 3 | Algorithms; Systems; | **Risks** principle in healthcare mainly refers to the patient's social and safety due to biased health AI systems |
| Privacy* | 1 | Data Access; Systems | **Privacy** is the protection of patients' critical information in the data |

**Table 2. Taxonomy/ Codes related to ethical principles, frequency, and relevant terms from the data,* Agreed codes from 3 coder**

**Results and Discussion**

The results of this study identify the major principles and challenges which will help the researchers to mold their healthcare AI design with proper ethical principles. In this research, E is used to refer to the experts of this study. Table 3 provides the overall query of all the words that appeared from the expert's data in a word query format. From the query, we can easily identify Justice, privacy, bias, and lack of regulation as the most potential principle towards ethical AI. The ethical principles found in the query are in line with the ethical principles identified from other works, which confirms the unified factors for the unified ethical framework.

| Technology Levels | Word | Length | Count | Weighted Percentage (%) |
|---|---|---|---|---|
| Data access; | privacy | 7 | 6 | 0.65 |
| Algorithms; | justice | 7 | 8 | 0.87 |
|  | biased | 6 | 5 | 0.54 |
|  | noninterpretable | 16 | 3 | 0.33 |
| Policy; Organizational; | regulations | 11 | 4 | 0.43 |
| Systems; | risk | 4 | 4 | 0.43 |

**Table 3. Word Query Results with Length, Count, and Weighted Percentage**

*Major Ethical Principles of AI in Healthcare at Different Technological Levels*

Our data analysis, queries, and data patterns made us reevaluate the Goirand et al., (2021)'s level categorization (ethical principle, design, technology, organizational, regulatory) of healthcare AI challenges and ethical principles.

**Data Access Level**

**Privacy**

Expert E6 stated that AI has positive effects, but there are risks related to privacy. Expert E6 mentions that the real privacy risk depends on organizations accessing health and education-related information. Expert E4 presented a scenario of how AI invades the user's privacy. Expert E21 mentions it is terrifying that AI will be developed as a guardian to look after us in the upcoming future. AI will soon be dictating what to do. When to get up? When to sleep, and other related things that invade user privacy. Expert E10 mentioned that if AI is used for targeted advertising, there is a risk of losing anonymity.

**Algorithms Level**

**Beneficence**

The study shows experts E4, E8, E12, E13, E16, E18, E24, E29, E31, E32, E34, and E36 have managed to discuss how AI has benefitted the healthcare domain. Besides the criticism AI faced, AI can positive impact such as in health care AI aid, and early detection of diseases.

Expert E18 discusses that AI and ML are most effective in the healthcare domain, the AI algorithm will examine more than a thousand possibilities in seconds. Expert E24 states AI in health and transportation will make difference in the lives of most people. AI has benefitted many through ML applications influencing speech recognition, language translation, search efficiency and effectiveness, and medical diagnosis. AI beneficence ranges too many dimensions thus expecting to improve quality of life.

**Bias**

Bias can be defined as systematic discrimination against a particular entity or a group of entities depending on their traits or characteristics. Experts E1, E10, E15, E19, E27, E29, E34, and E36 stated their opinion of bias foremost in healthcare AI. AI systems are specially used for the decision-making process, since the training data of AI algorithms are mostly biased, there might be scenarios where people force their own opinions on the AI systems to produce a biased result. According to expert E36, bias is an already existing issue in AI systems that can have an adverse effect on people's lives. In addition to that implicit bias is hard to ignore because it is difficult to identify.

**Justice and Solidarity**

Experts are concerned about the dangerous impacts of unequal non-transparent AI algorithms. Experts E2, E11, E14, E20, and E33 have suggested ethical principle implementation for bringing equality in AI.

**Interpretability**

Opacity, transparency, and explanation may have different meanings in different fields of study but generally, they have the same meaning in the ML ecosystem. Expert E32 mentions that there is a need for us to worry about the AI being used to shape the policy in nontransparent, noninterpretable, and nonreproducible ways. Expert E33 expressed concerns regarding transparency issues with AI Moreover, Expert E37 predicts that there would be changes in ethical principles regarding AI, but they may not be convincing enough or transparent.

**Systems Level**

**Accountability**

Accountability issues occur due to the low performance of AI. Accountability in AI ranges in various domains alongside health. Experts E3, E6, E10, E15, E18, and E33 have raised concerns about how AI is perceived when accountability comes into play. According to expert E10, one of the accountability issues of the AI algorithm was identified under the context of color discrimination, for example, people with darker skin tones receiving high doses of radiation to penetrate the pigment in the skin for developing a clearer medical image for treatment can result in serious repercussions on the health of the person.

**Reliability**

Reliability deals with the efficiency of implementing AI. Expert E19 discusses a few accuracy problems related to facial recognition software. Accuracy problems are more visible recognizing 'nonwhite faces' leading to reliability issues related to AI algorithms.

According to expert E28, human decision-making makes the AI system more reliable. Human decision-making consists of designing the algorithm and deciding what data to include and exclude in the dataset. Expert E11 adds that AI systems need to be monitored to make sure humans make the final decision.

**Risks**

Wrong and unethical use of AI in healthcare can pose patient safety risks. Experts E17, E23, E28, and E31 have expressed their concern over the patient safety risk due to the unethical use of AI in healthcare. In this context, E28 kept safety risks alongside privacy and security risk.

**Policy and Organizational Level**

**Lack of proper regulation**

Nine experts E2, E7, E12, E15, E17, E20, E31, E33, and E34 among 36 experts expressed the need for regulatory frameworks and reporting guidelines for AI in healthcare. Many experts believe that reporting guidelines will help in building ethical AI. In addition, E34 mentions the importance of adherence to reporting guidelines while designing AI.

## A Unified Utilitarian Ethics Framework

In medicine and healthcare, rule-based utilitarianism is considered better than act-based utilitarianism as preformed rules based on evidence assist in better decision making and there is no prediction or calculation of harm[11]. Similarly, when decision-making is assisted by technology, AI technology is developed from evidence derived from previous health record data. In our proposed model, we are proposing a utilitarian approach for designing an ethical framework with the influence of variables derived from our study at different technology levels. Potential variables are beneficence, justice, bias, interpretability, reliability, risks, privacy, and accountability. Moreover, our framework consists of the "lack of regulation" principle at the organizational policy level. In AI4people, the ethical framework consisted of autonomy, beneficence, non-maleficence, justice, and a new variable explicability which is a representation of "accountability" meaning how AI can be responsible for the work. The goal of the AI4people ethical framework was to build a good AI society. However, challenges faced due to the appearance of new AI technologies and the frequent use of AI in healthcare have moved researchers' focus toward some new important principles in the ethical framework. Usually, in utilitarianism, the consequence of the actions should bring maximized happiness. These consequences of actions are influenced by ethical principles and require reporting guidelines. Our proposed ethical framework (shown in Figure 2.) comprised important ethical principles as an influencer of actions at different technology levels in the utilitarianism approach for generating better healthcare outcomes.

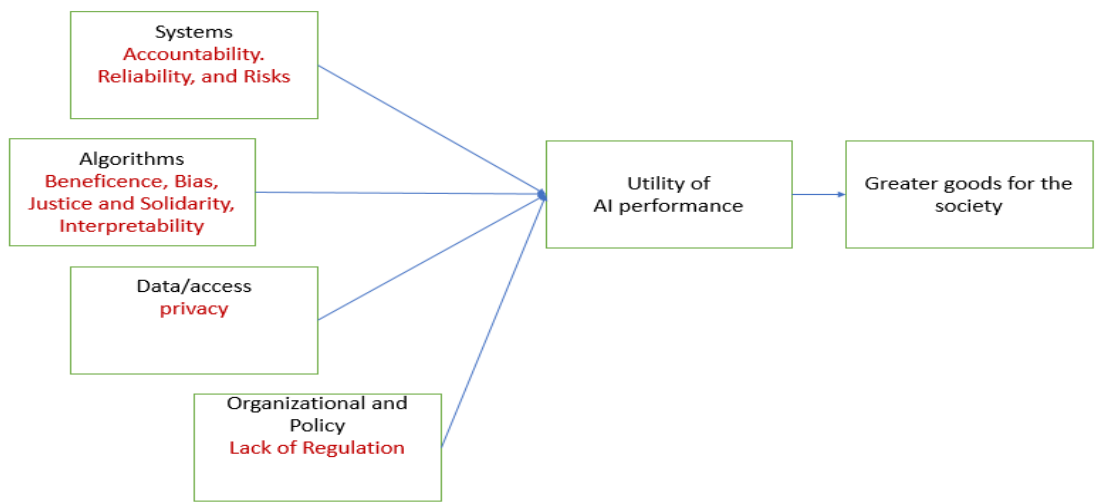

**Figure 2. Utilitarian Approach-based Ethical Framework**

**Theory and Practical Implications**

Ethics is usually misrepresented as ethical framework varies from domain to domain. Researchers usually mix different ethical principles and their application levels. Some researchers have indicated that ethical principles should be implemented at different levels. However, application at the technology level has not been demonstrated deeply as ethical principles of data access level, algorithms level, and systems level are quite different. We provided a broader view of the technology-level application of ethical principles in our study. However, our study has some limitations as we only conducted an analysis of small sample-sized data with only 3 coders. There were some principles such as autonomy which appeared in the word query but were not addressed by the coders. A more robust study can be conducted on the ethical AI framework by adding a survey study to quantify the results in the future.

In order to have a practical implementation of AI in healthcare, we divided the technology level into data access level, algorithm level, and systems level [12]. A healthcare system is the hardware and software infrastructure which is used to deploy the applications containing proper usability [15]. Algorithms can be referred to as model that learns from data patterns to make predictions. The systems level contains risks, accountability, and reliability as system transparency can bring accountability and reduce risks [13]. Similarly, there is a need for transparency in the algorithm which can be implemented through interpretability and fairness, then the algorithm will be more useful in healthcare. Therefore, the algorithms level consists of beneficence, bias, interpretability, justice & solidarity, and the data access level contains privacy [14]. In addition, we categorized the "lack of regulation" principle at the policy and organizational level.

Practical implications of the proposed framework can be recently developed ChatGPT or GPT-based applications. We can observe different news on the unethical use of AI in recent days such as 'Mind-reading AI' developed by Japan's Osaka University, Samsung's ChatGPT data leaking incident, and warning of "the Godfather of AI" Geoffrey Hinton regarding unethical AI use. Although, Google is hiring ethicists and philosophers to make AI more on the moral ground. Technological reforms in data access, systems, and algorithms based on utilitarianism ethics can bring ethical aspects to future ChatGPT which will restrict it from unethical use and will enable ChatGPT to be used for good purposes only.

**Conclusion**

The overall finding of this study suggests that AI experts are concerned about the successful ethical AI design in healthcare. Designing an ethics-infused framework of AI by mitigating problematic issues such as privacy issues, misuse of data, and interpretability can only result in greater trustworthiness of the system and an increase in the use of AI in the healthcare domain by clinicians, physicians, healthcare professionals, and other stakeholders. Adherence to the reporting guidelines can be the steppingstone toward the successful design of ethical AI, which can eventually lead to the actual use of AI in healthcare. Our study will pave the way to encourage future researchers to dive deeply into the ethical challenges presented by AI and find more efficient solutions to resolve them. Future work will provide examples of actions that fulfill the conditions of the approach based on the proposed framework.